\documentclass[12pt]{article}
\usepackage[dvips]{color}
\usepackage{epsfig}
\usepackage{amsmath}
\usepackage{graphicx}
\usepackage{cite}
\usepackage{subcaption}

\begin{document}

\title{\bf GUP-Corrected van der Waals Black Holes}
\author{{ \"{O}zg\"{u}r \"{O}kc\"{u} \thanks{Email: ozgur.okcu@ogr.iu.edu.tr}\hspace{1mm},
Ekrem Aydiner \thanks{Email: ekrem.aydiner@istanbul.edu.tr}} \\
{\small {\em Department of
		Physics, Faculty of Science, Istanbul University,
		}}\\{\small {\em Istanbul, 34134, Turkey}} }
\maketitle

\begin{abstract}
In this paper, we study the generalized uncertainty principle (GUP) effects for the van der Waals (vdW) black holes. In order to obtain the GUP-corrected solution, we consider GUP corrected black hole temperature. We also study the thermodynamics and phase transition
of GUP-corrected vdW black holes. We compare the differences between thermodynamic properties of both modified and original solutions. We show that P-V criticality is physically acceptable in the presence of GUP-correction.\\

\end{abstract}

\section{Introduction}
\label{intro}
Since the revolutionary papers of Bekenstein and Hawking, black hole thermodynamics has been one of the most important subjects among the researchers in the scientific community \cite{Bekenstein1972,Bekenstein1973,Bardeen1973,Hawking1974,Bekenstein1974,Hawking1975}. Considering black holes with temperature and entropy gives us new opportunities to explore many interesting thermodynamic phenomena. Furthermore, black hole thermodynamics gives the first hints of quantum gravity. It also gives the fundamental links between general relativity, thermodynamics and quantum mechanics. One may naturally ask whether black hole as a thermodynamic system  shares any similarities with general thermodynamic system or not. These similarities become more clear and certain for black holes in anti-de Sitter (AdS) spacetime. 

Black holes in AdS spacetime have been studied widely in the literature since the pioneer paper of Hawking and Page \cite{Hawking1983}. They found a first order phase transition between Schwarzschild AdS black hole and thermal AdS space. Interestingly when Schwarzschild AdS black hole is generalized to charged or rotating case, it shows the vdW fluids like behaviours. The authors of \cite{Chamblin1999a,Chamblin1999b} studied the thermodynamics of charged AdS black holes and they found vdW like first order small$-$large black hole phase transition. This type of phase transition becomes more clear in the extended phase space where the cosmological constant is considered as a thermodynamic pressure. Once treating the cosmological constant as a thermodynamic pressure,
\begin{equation}
\label{pressure}
P=-\frac{\Lambda}{8\pi}\,,
\end{equation}
naturally gives its conjugate quantity as a thermodynamic volume\footnote{Based on this idea, Kastor et al.\cite{Kastor2009} showed that mass of AdS black hole is identified with enthalpy of spacetime.}
\begin{equation}
\label{thermoVolume}
V=\left(\frac{\partial M}{\partial P}\right)_{S,Q,J}\,.
\end{equation}
The charged AdS black hole thermodynamics and phase transition were studied by Kubiznak and Mann\cite{Kubiznak2012}. They showed the charged AdS first order small$-$large  black hole phase transition has the same characteristic behaviour with vdW fluids. They were also obtained critical exponents which coincide the exponents of vdW fluids. Up to now, it can be seen from literature that their study has been extended for various solutions of black holes in the AdS spacetime \cite{Gunasekaran2012,Spallucci2013,Zhao2013,Behlaj2013,Cai2013,Mo2014,Xu2014,Li2014,Ma2014,Belhaj2015,Kubiznak2015,Hennigar2015,Caceres2015,Wei2016,Pourhassan2016,Hendi2016,Momeni2017,Ovgun2018,Sun2018,Jamil2018,Nam2018a,Nam2018b,Kuang2018,Zhang2018,Okcu2017,Okcu2018,Mo2018,Zhao2018,Yekta2019}\footnote{One can refer to the recent comprehensive reviews \cite{Altamirano2014,Kubiznak2017} and references therein.}.

Based on the above fact, Rajagopal et al. \cite{Rajagopal2014} obtained vdW black hole solution which has the same thermodynamics with vdW fluids\footnote{See \cite{Pradhan2016,Hu2017} for the thermodynamics of vdW black holes.}. In their interesting paper, they also argued the corresponding stress energy tensor for their solution. They found that the stress energy tensor obeys energy conditions for a certain range of metric parameter. They also showed that their solution is interpreted as near horizon metric. Following the methods in \cite{Rajagopal2014}, a few AdS black hole solutions, which match the thermodynamics of a certain equation of states, were proposed in the literature. In \cite{Delsate2015}, Delsate and Mann generalized the vdW solution in higher dimensions. In \cite{Setare2015}, Setare and Adami obtained Polytropic black hole solution which has the identical thermodynamics with that of the polytropic gas. Interestingly, under the small effective pressure limit, Abchouyeh et al. \cite{Abchouyeh2017} obtained Anyon black hole solution which corresponds to thermodynamics of Anyon vdW fluids. Anyons are the particles that have the intermediate statistics between Fermi-Dirac and Bose-Einstein statistics. Then exact solution of Anyon black hole \cite{Xu2018} was obtained by Xu. Finally, Debnath constructed a black hole solution whose thermodynamics matches the thermodynamics of modified Chaplygin gas \cite{Debnath2018}.

Furthermore, black hole thermodynamics is considered in the context of quantum gravity since the effects of quantum gravity are no longer negligible near the Planck scale. Therefore, thermodynamics of black hole was modified in various quantum gravity approaches \cite{Govindarajan2001,Mann1998,Sen2013,Das2002,Feng2017,Feng2019}. On the other hand, black holes as a gravitational system may give us information about nature of the quantum gravity. Motivated by this fact, Upadhyay and Pourhassan \cite{Upadhyay2019} studied the modification of higher dimensional vdW black hole for the thermal fluctuations interpreted as the quantum effects. They also investigated the thermal fluctuations effects on the thermodynamics of higher dimensional vdW black hole. 

It is well known that GUP is one of the phenomenological quantum gravity models and is considered as an modification of standard uncertainty principle \cite{Maggiore1993,Kempf1995,Nozari2012a}. Therefore, it is possible to modify the thermodynamics properties of black hole by taking into account the GUP effects near the Planck scale\cite{Adler2001,Nozari2005,Nozari2008,Nowakowski2009a,Nowakowski2009b,Banerjee2010,Nozari2012b,Ali2012,Gangopadhyay2014,Abbasvandi2016,Feng2016,Luciano2019,Villalpando2019,Sakalli2016,Kanzi2019,Jusufi2020,Bosso2020,Gecim2020,Xiang2009}.
So far, to our best knowledge, GUP has never been extended to the thermodynamics of four dimensional vdW black holes. In this study, we want to explore GUP effects for four dimensional vdW black holes.

The paper is arranged as follows: We first review the heuristic derivation of GUP modified Hawking temperature which was proposed by Xiang and Wen \cite{Xiang2009}. Next, we modify the vdW solution by using the modified Hawking temperature and then we check the energy conditions for stress energy tensor in Sect. \ref{GUPBH}. In Sect. \ref{GUPBHT}, we investigate the GUP corrected thermodynamics quantities and phase transition. Finally, we discuss our results in Sect. \ref{Concl.}. (We use the units $G_{N}=k_{B}=c=L_{pl}=1$).
\section{GUP-Corrected Black Hole Temperature}
\label{GUPR}
Here, we will briefly review a generic GUP correction approach to semi-classical Hawking Temperature \cite{Xiang2009}. The simplest form of GUP is given by \cite{Maggiore1993}
\begin{equation}
\Delta x\Delta p\geq\hbar+\frac{\alpha}{\hbar}\Delta p^{2}\,,
\label{GUP}
\end{equation}
where $\alpha$ is a positive constant. To get the correction to the black hole thermodynamics, we need to solve this inequality for the momentum uncertainty $\Delta p$. The solution of (\ref{GUP}) is given by
\begin{equation}
\label{deltaP1}
\frac{\hbar}{2\alpha}\left(\Delta x+\sqrt{\Delta x^{2}-4\alpha}\right)\geq\Delta p\geq\frac{\hbar}{2\alpha}\left(\Delta x-\sqrt{\Delta x^{2}-4\alpha}\right)\,,
\end{equation}
where we choose the lower bound of the inequality since we can recover the standard uncertainty principle in the limit $\alpha\rightarrow0$. Series expansion of (\ref{deltaP1}) yields
\begin{equation}
\label{expandedDeltaP}
\Delta p\geq\frac{\hbar}{\Delta x}+\frac{\hbar\alpha}{\Delta x^{3}}+\mathcal{O}(\alpha^{2})\,.
\end{equation}
Therefore, by using the above statement we can write $\Delta x\Delta p$ as
\begin{equation}
\label{effectivePlanck}
\Delta x\Delta p\geq\hbar\left(1+\frac{\alpha}{\Delta x^{2}}+\mathcal{O}(\alpha^{2})\right)=\hbar_{eff}\,,
\end{equation}
where the rhs of the inequality can be considered as an effective Planck constant $\hbar_{eff}$. On the other hand, the smallest increase of area for the black hole, which absorbs a particle, is given by
\begin{equation}
\label{areaChange}
\Delta A\geq\Delta x\Delta p\,.
\end{equation}
Taking the uncertainty of position $\Delta x\approx2r_{h}$ and using the (\ref{effectivePlanck}) with the (\ref{areaChange}), one can obtain the increase of the area as
\begin{equation}
\label{areaChange2}
\Delta A\geq\gamma\hbar_{eff}=\gamma\hbar\left(1+\frac{\alpha}{4r_{h}^{2}}+\mathcal{O}(\alpha^{2})\right)\, ,
\end{equation}
where $r_{h}$ stands for event horizon of black hole, and $\gamma$ is a calibration factor which can be obtained in the limit $\alpha\rightarrow0$. Moreover, when the particle is absorbed, the minimum increase in black hole entropy is given $(\Delta S)_{min}=\ln2$. So we can obtain
\begin{equation}
\label{dA/dS}
\frac{dA}{dS}\simeq\frac{(\Delta A)_{min}}{(\Delta S)_{min}}=\frac{\hbar\gamma}{\ln2}\left(1+\frac{\alpha}{4r_{h}^{2}}+\mathcal{O}(\alpha^{2})\right)\,.
\end{equation}
Using the temperature of black hole $T=\frac{dA}{dS}\times\frac{\kappa}{8\pi}$ with (\ref{dA/dS}), we finally find the GUP corrected temperature
\begin{equation}
\label{GUPTemp}
T=\frac{\hbar\gamma}{\ln2}\left(1+\frac{\alpha}{4r_{h}^{2}}+\mathcal{O}(\alpha^{2})\right)\times\frac{\kappa}{8\pi}\,,
\end{equation}
where $\kappa=f'(r_{h})/2$ is the surface gravity of the black hole, and prime denotes the derivative with respect to r. In order to find $\gamma$, we should check the GUP-modified temperature in the limit $\alpha\rightarrow0$. The (\ref{GUPTemp}) should give the standard result $T=\frac{\hbar\kappa}{2\pi}$ when $\alpha$ goes to zero. As a result, we find the calibration factor $\gamma=4\ln2$. We find the GUP modified temperature 
\begin{equation}
\label{GUPTemp2}
T=\frac{\hbar_{eff}\kappa}{2\pi}=\frac{\hbar\kappa}{2\pi}\left(1+\frac{\alpha}{4r_{h}^{2}}\right)\,,
\end{equation}
for static and spherically symmetric black holes.

In this section, we briefly review the generic GUP correction to black hole temperature. In the next section, we will use (\ref{GUPTemp2}) to modify the vdW black hole solution. For the clarity of the following discussions, we choose $\hbar=1$ in the rest of the paper.

\section{GUP-Corrected vdW Black Holes}
\label{GUPBH}
It is well-known that vdW equation of state is a generalized version of ideal gas equation. It is given by \cite{Johnston2014}
\begin{equation}\label{vdW}
P=\frac{T}{v-b}-\frac{a}{v^{2}} \,,
\end{equation}
where $v=V/N$ is the specific volume, $a>0$ constant is a measure of the attraction between the particles, and $b>0$ is a measure of the particle volume.  One can use vdW equation for describing the behaviour of liquid$-$gas phase transition. In order to construct a black hole solution whose thermodynamic matches with that of (\ref{vdW}), we start with the following spherically symmetric  ansatz for the metric
\begin{equation}\label{metric1}
ds^{2}=-f(r)dt^{2}+\frac{dr^{2}}{f(r)}+r^{2}d\Omega^{2} \,,
\end{equation}
\begin{equation}\label{metric2}
f=\frac{r^{2}}{l^{2}}-\frac{2M}{r}-h(r,P) \,,
\end{equation}
where $l$ is the AdS radius, and the function $h(r,P)$ can be obtained from GUP-corrected black hole temperature. Now, we assume that the given metric is a solution of the Einstein field equations, $G_{\mu\nu}+\Lambda g_{\mu\nu}=8\pi T_{\mu\nu}$. In order to proceed discussion, we choose the stress energy tensor $T^{\mu\nu}$ as an anisotropic fluid source in the following form
\begin{equation}\label{stressEnergy}
T^{\mu\nu}=\rho e_{0}^{\mu}e_{0}^{\nu}+\sum_{i}p_{i}e_{i}^{\mu}e_{i}^{\nu} \,,
\end{equation}
where $e_{i}^{\mu}$, $\rho$ and $p_{i}$ denote the components of the vielbein ($i=1,2,3$), energy density and principle pressure, respectively. We need a physically meaningful stress energy source for our metric ansatz. Therefore, we require our corresponding stress energy tensor should satisfy certain energy conditions such as weak, strong and dominant energy conditions. We will consider these conditions on the energy density $\rho$ and principal pressures $p_{i}$ after determining the metric. Using Einstein field equations with the stress energy tensor in  (\ref{stressEnergy}), $\rho$ and $p_{i}$ are given by
\begin{equation}\label{energyDensityWithp1}
\rho=-p_{1}=\frac{1-f-rf'}{8\pi r^{2}}+P \,,
\end{equation}
\begin{equation}\label{principalPressure23}
p_{2}=p_{3}=\frac{rf''+2f'}{16\pi r}-P\,,
\end{equation}
where the relation between thermodynamic pressure and the AdS radius is defined by
\begin{equation}
\label{pressure2}
P=-\frac{\Lambda}{8\pi}=\frac{3}{8\pi l^{2}}\, .
\end{equation}
 On the other hand, the mass of black hole can be obtained from $f(r_{h})=0$
\begin{equation}\label{mass}
M=\frac{4}{3}\pi r_{h}^{3}P-\frac{h(r_{h},P)r_{h}}{2}\,.
\end{equation}
At this point, we can give GUP corrected thermodynamic quantities
\begin{equation}\label{temperature}
T=\frac{\hbar_{eff}\kappa}{2\pi}=\left(1+\frac{\alpha}{4r_{h}^{2}}\right)\left(2Pr_{h}-\frac{h(r_{h},P)}{4\pi r_{h}}-\frac{1}{4\pi}\frac{\partial h(r_{h},P)}{\partial r_{h}}\right)\,,
\end{equation}
\begin{equation}\label{entropy}
S=\int\frac{dM}{T}=\pi r_{h}^{2}-\frac{\alpha\pi}{4}\ln\left(\frac{4r_{h}^{2}+\alpha}{\alpha}\right)\,,
\end{equation}
\begin{equation}\label{volume}
V=\left(\frac{\partial M}{\partial P}\right)=\frac{4}{3}\pi r_{h}^{3}-\frac{r_{h}}{2}\frac{\partial h(r_{h},P)}{\partial P}\,,
\end{equation}
\begin{equation}
\label{heatCapacity}
C_{P}=\frac{\partial M/\partial r_{h}}{\partial T/\partial r_{h}}=\frac{8\pi r_{h}^{4}\left(-h+r_{h}\left(\frac{3r_{h}}{l^{2}}-\frac{\partial h}{\partial r_{h}}\right)\right)}{(4r_{h}^{2}+3\alpha)h+r_{h}\left((4r_{h}^{2}-\alpha)\left(\frac{3r_{h}}{l^{2}}-\frac{\partial h}{\partial r_{h}}\right)-r_{h}(4r^{2}+\alpha)\frac{\partial^{2}h}{\partial r_{h}^{2}}\right)}\,,
\end{equation}
where we choose integration constant as $\frac{\alpha\pi}{4}\ln(\alpha)$ to make a dimensionless logarithmic term in (\ref{entropy}). One can also define the specific volume of black hole by \cite{Altamirano2014}
\begin{equation}
\label{specificVolume}
v=\frac{V}{N}\,,
\end{equation}
where $N=\frac{4(d-1)}{d-2}\frac{\mathcal{A}}{L^{2}_{pl}}$ is proportional to the horizon area of black hole $\mathcal{A}$. In $d=4$ dimensions, the specific volume is given by
\begin{equation}
\label{specificVolume2}
v=\frac{3}{2\pi r_{h}^{2}}\left[\frac{4}{3}\pi r^{3}-\frac{r_{h}}{2}\frac{\partial h(r_{h},P)}{\partial P}\right]\,.
\end{equation}

In order to construct a metric solution whose thermodynamics is identical with the vdW fluids, we assume that $h(r,P)=A(r)-PB(r)$. First, we obtain the temperature from (\ref{vdW})
\begin{equation}\label{vdW2}
T=\left(P+\frac{a}{v^{2}}\right)(v-b) \,,
\end{equation}
and using the equality between (\ref{temperature}) and (\ref{vdW2}), we can write 
\begin{equation}
\label{equation}
\left(1+\frac{\alpha}{4r_{h}^{2}}\right)\left(2Pr_{h}-\frac{h}{4\pi r_{h}}-\frac{h'}{4\pi}\right)=\left(P+\frac{a}{v^{2}}\right)(v-b)\,,
\end{equation}
where $v=2r_{h}+\frac{3B}{4\pi r_{h}}$. The (\ref{equation}) is organized in the form $F_{1}(r)+F_{2}(r)P=0$, where the functions $F_{1}(r)$ and $F_{2}(r)$ depend on the functions $A(r)$ and $B(r)$ and their derivatives. With ansatz $h$, we get two ordinary differential equations from (\ref{equation}). In other words, we should independently set the functions $F_{1}(r)$ and $F_{2}(r)$ equal zero,
\begin{equation}
\label{F1}
F_{1}(r)=-\frac{1}{4\pi}\left(1+\frac{\alpha}{4r^{2}}\right)\left(A'+\frac{A}{r}\right)-\frac{16\pi^{2}ar^{2}}{(8\pi r^{2}+3B)^{2}}\left(2r+\frac{3B}{4\pi r}-b\right)=0\,,
\end{equation}
\begin{equation}
\label{F2}
F_{2}(r)=\frac{1}{4\pi}\left(1+\frac{\alpha}{4r^{2}}\right)\left(B'+\frac{B}{r_{h}}\right)-\frac{3B}{4\pi r}+\frac{\alpha}{2r}+b=0\,.
\end{equation}
\begin{figure}
	\centerline{\includegraphics[width=10cm]{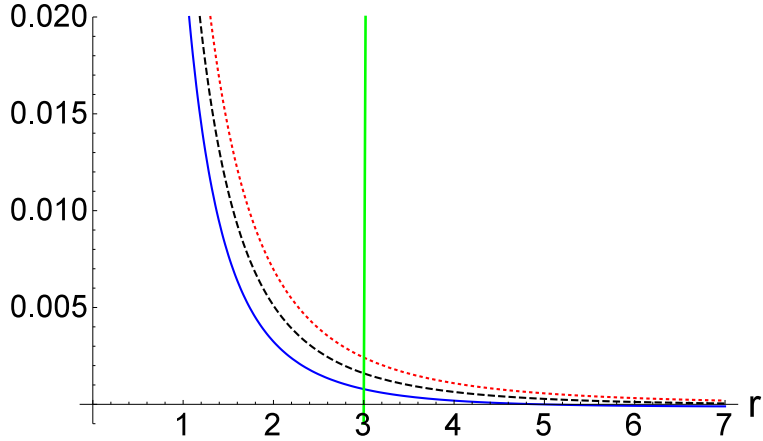}}
	\vspace*{8pt}
	\caption{(Colour online). We only show the energy density $\rho$ ( black dashed line), $\rho+p_{2}$ (red dotted line) and $\rho-p_{2}$ (blue solid line) for the sufficiently small pressure. Thick green vertical line presents the event horizon.  We set $a=\frac{1}{2\pi}$, $b=1$, $P=0.001$ and $M=0.16$ \label{f1}}
\end{figure}
The coefficient $B(r)$ is obtained from (\ref{F2})
\begin{equation}
\label{Br}
B(r)=4\pi br-\frac{8\pi r^{2}}{3}+8\epsilon r^{2}+\left(\frac{2\pi b}{3r}+3\epsilon\right)\alpha+\mathcal{O}(\alpha^{2})\,.
\end{equation}
For simplicity, we choose integration constant as $\epsilon=\pi/3$. Hence, we obtain 
\begin{equation}
\label{Br2}
B(r)=4\pi br+\left(1+\frac{2b}{3r}\right)\pi\alpha\,.
\end{equation}
Substituting (\ref{Br2}) into (\ref{F1}) and then expanding $F_{1}(r)$ up to second order of $\alpha$, we can find the following equation: 
\begin{equation}
\label{F1new}
F_{1}(r)=-\frac{1}{4\pi}\left(1+\frac{\alpha}{4r^{2}}\right)\left(A'+\frac{A}{r}\right)-\frac{2a(b+r)}{(3b+2r)^{2}}+\frac{a(b+2r)(2b+3r)}{4r^{2}(3b+2r)^{3}}\alpha=0\,,
\end{equation}
and this equation yields the solution
\begin{eqnarray}
\label{Ar}
&A(r)=-2\pi a+\frac{\pi ab^{2}(243b^{4}(3b+2r)-9b^{2}(18b+17r)\alpha-(57b+43r)\alpha^{2})}{r(3b+2r)^{2}(9b^{2}+\alpha)^{2}}\nonumber\\+&\frac{a\pi\alpha^{3/2}(-27b^{^{4}}+2\alpha b^{2}+5\alpha^{2})\arctan(2r/\sqrt{\alpha})}{2r(9b^{2}+\alpha)^{3}}+\frac{2\pi ab(1458b^{6}+378\alpha b^{4}+9\alpha^{2}b^{2}-5\alpha^{3})}{r(9b^{2}+\alpha)^{3}}\ln\left(\frac{3b+2r}{2b}\right)\nonumber\\&+\frac{\pi\alpha ab(108b^{4}+45\alpha b^{2}+7\alpha^{2})}{r(9b^{2}+\alpha)^{3}}\ln\left(\frac{4r^{2}+\alpha}{4b^{2}}\right)\,,
\end{eqnarray}
where we choose the suitable integration constant to obtain the dimensionless logarithmic terms. Using the solutions in (\ref{Br2}) and (\ref{Ar}), we determine the modified $h(r,P)$ functions and therefore we get the GUP-corrected vdW black hole solution. As it can be seen from (\ref{Br2}) and (\ref{Ar}), the result in \cite{Rajagopal2014} are obtained in the limit $\alpha\rightarrow0$.
\begin{figure}
	\centerline{\includegraphics[width=10cm]{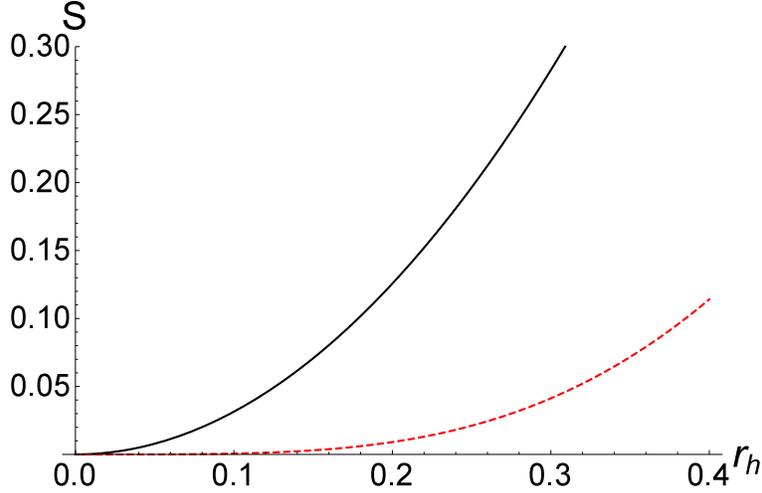}}
	\vspace*{8pt}
	\caption{(Colour online). Semi-classical (black solid line) and GUP-corrected (red dashed line) entropies as function of $r_{h}$. We set $a=\frac{1}{2\pi}$, $b=1$.\label{f2}}
\end{figure}

For a valid physical solution, stress energy tensor should satisfy the certain energy conditions. Therefore, the energy conditions should be checked for finding a physically meaningful solution \cite{Poisson2004}:
\begin{eqnarray}
Weak:&\qquad  \rho \geq 0\ ,\         \rho+{p}_i \geq 0\\
Strong:&\quad  \rho+{\sum}_i{ p}_i\geq 0\,  \qquad   \rho+{p}_i \geq 0\\
Dominant:&\qquad  \rho \geq 0\ , \qquad \rho \geq |{p}_i|\, .
\end{eqnarray}
In Fig.(\ref{f1}), we check the energy conditions. It is possible to satisfy the all energy conditions near the outer of the horizon for the sufficiently small pressure. Since $p_{2}$ is positive for a sufficiently large $r$, it is not displayed in the figure. It seems our solution is physically valid near horizon.

In the next section, we will investigate the thermodynamics of GUP-corrected vdW black holes. 

\section{Thermodynamics and Phase Transition}
\label{GUPBHT}
In the previous section, we obtained the GUP-corrected $h(r_{h},P)$ function. Therefore, we can explore the GUP effects for the thermodynamic quantities of vdW black hole. In Fig.(\ref{f2}), we show both semi-classical and GUP-corrected entropies of vdW black hole. It is clearly seen that GUP-corrected entropy is always smaller than semi-classical entropy. Moreover, modified entropy is a monotonically increasing function of $r_{h}$ in the region ($0<r_{h}<\infty$) since
\begin{equation}
\label{derivativeOfEntropy}
\frac{dS}{dr_{h}}=\frac{8\pi r_{h}^{3}}{\alpha+4r_{h}^{2}}\,.
\end{equation}

\begin{figure}
	\includegraphics[width=0.5\linewidth]{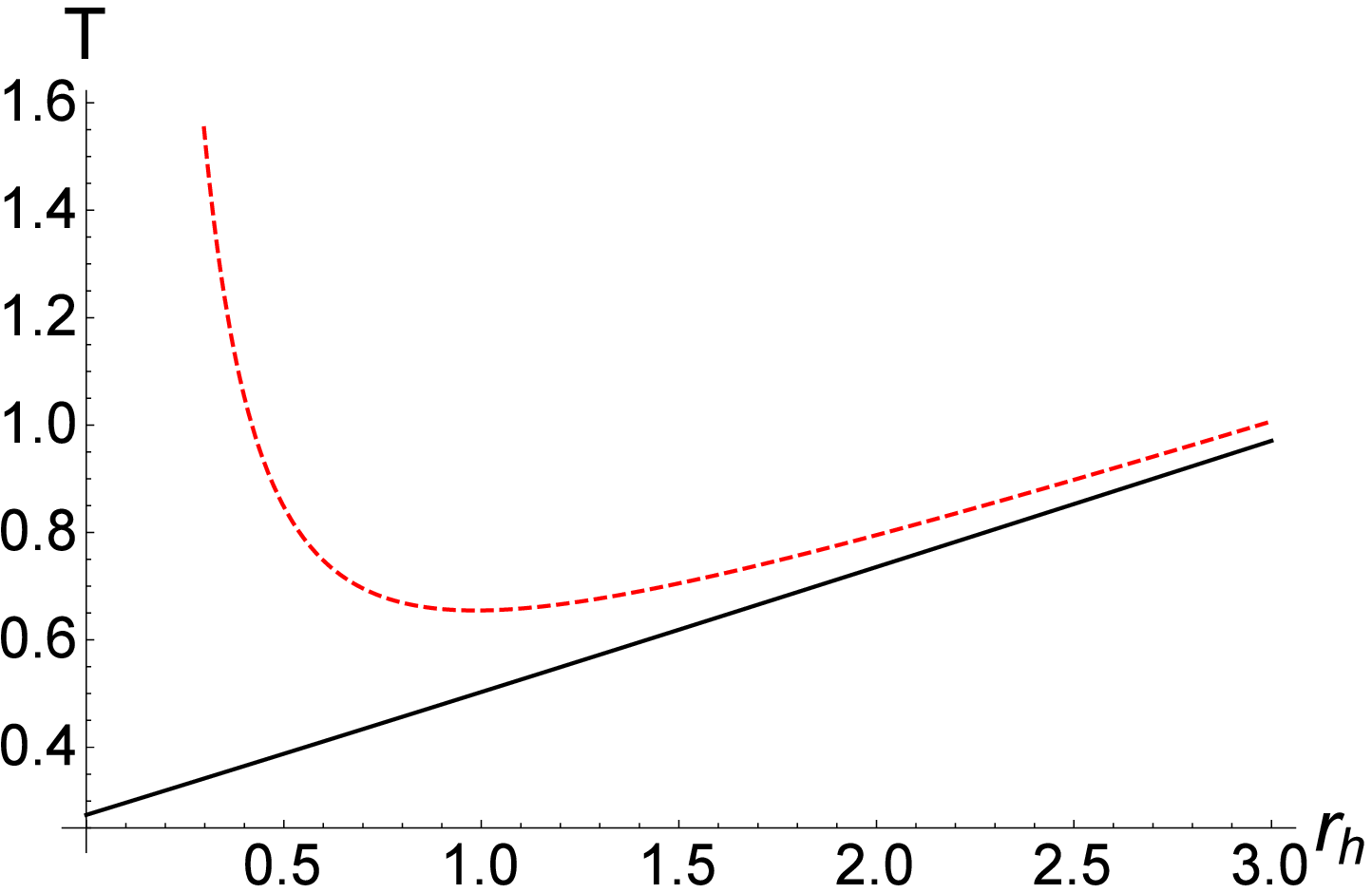}\hfil
	\includegraphics[width=0.5\linewidth]{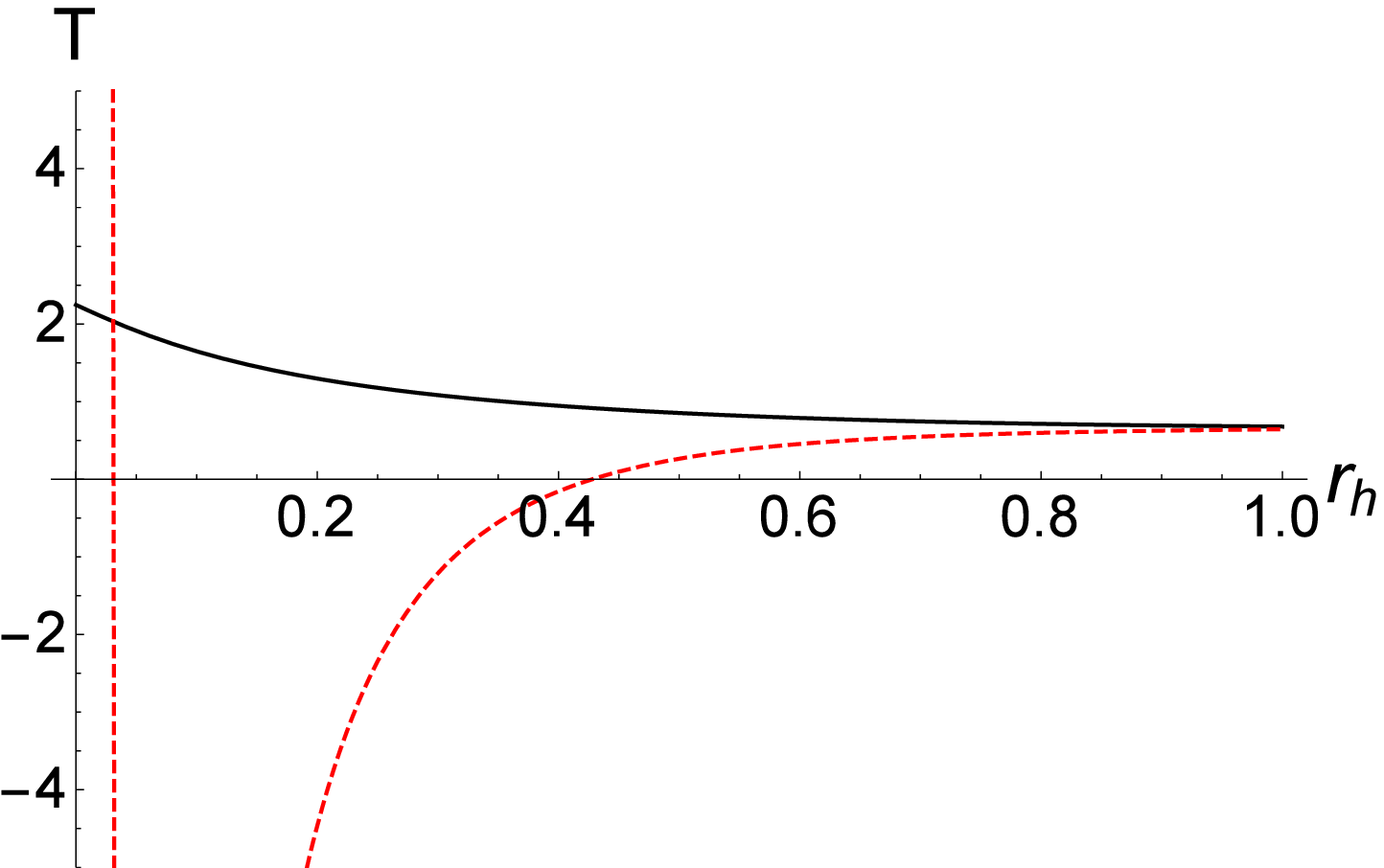}
	\caption{(Colour online). Semi-classical (black solid line) and GUP-corrected (red dashed line) temperatures as function of $r_{h}$. We set (a) $a=\frac{1}{2\pi}$, $b=l=1$. (b) $b=0.1$, $a=l=1$.}
	\label{f3}
\end{figure}

Corrected temperature may be larger or smaller than original temperature. In Fig.(\ref{f3}), semi-classical and modified temperatures are plotted in terms of $r_{h}$ for the suitable choices of parameters. In Fig.(\ref{f3}a), GUP-corrected temperature has an unstable branch for small black holes, while it has the same characteristic behaviour with semi-classical temperature due to negligible quantum gravity effects for the larger event horizons. The unstable branch corresponds to negative specific heat region. In this case, vdW black hole is thermodynamically unstable for the values of smaller event horizon in the presence of GUP effects. It is also clear that the temperature increases for the GUP correction. In Fig.(\ref{f3}b), corrected temperature shows an unstable branch for smaller event horizons and has ill-defined negative value behaviours for some event horizons. It is obvious that corrected temperature has a smaller value than original temperature for  certain event horizons and again has the same characteristic behaviour for larger event horizons due to negligible quantum gravity effects. One may consider difference between corrected and original temperature by
\begin{equation}
\label{tempDiff}
\Delta T=\frac{\alpha(12r_{h}^{2}+8br_{h}+\alpha)}{16r_{h}^{3}}-\frac{\alpha ar_{h}(b+2r_{h})(2b+3r_{h})}{4r_{h}^{3}(3b+2r_{h})^{3}}\,.
\end{equation}
When $\Delta T$ is positive (negative), corrected temperature is larger (smaller) than original temperature. The sign of (\ref{tempDiff}) is determined by the competition between first and second terms. For example, first term may give the most contributions for sufficiently larger event horizon. Therefore, corrected temperature may be larger than original temperature.

\begin{figure}
	\centering
	\includegraphics[width=0.5\linewidth]{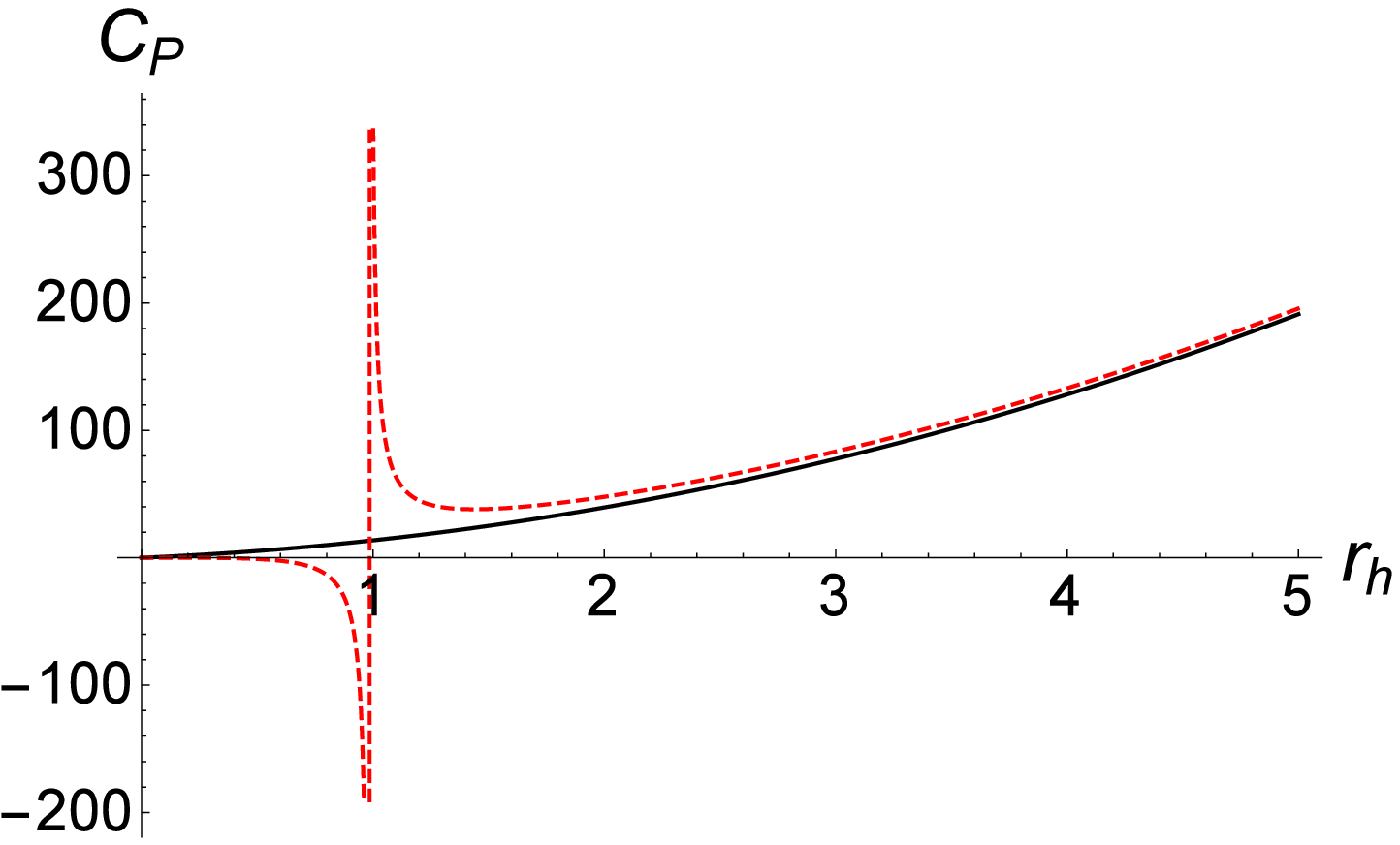}\hfil
	\includegraphics[width=0.5\linewidth]{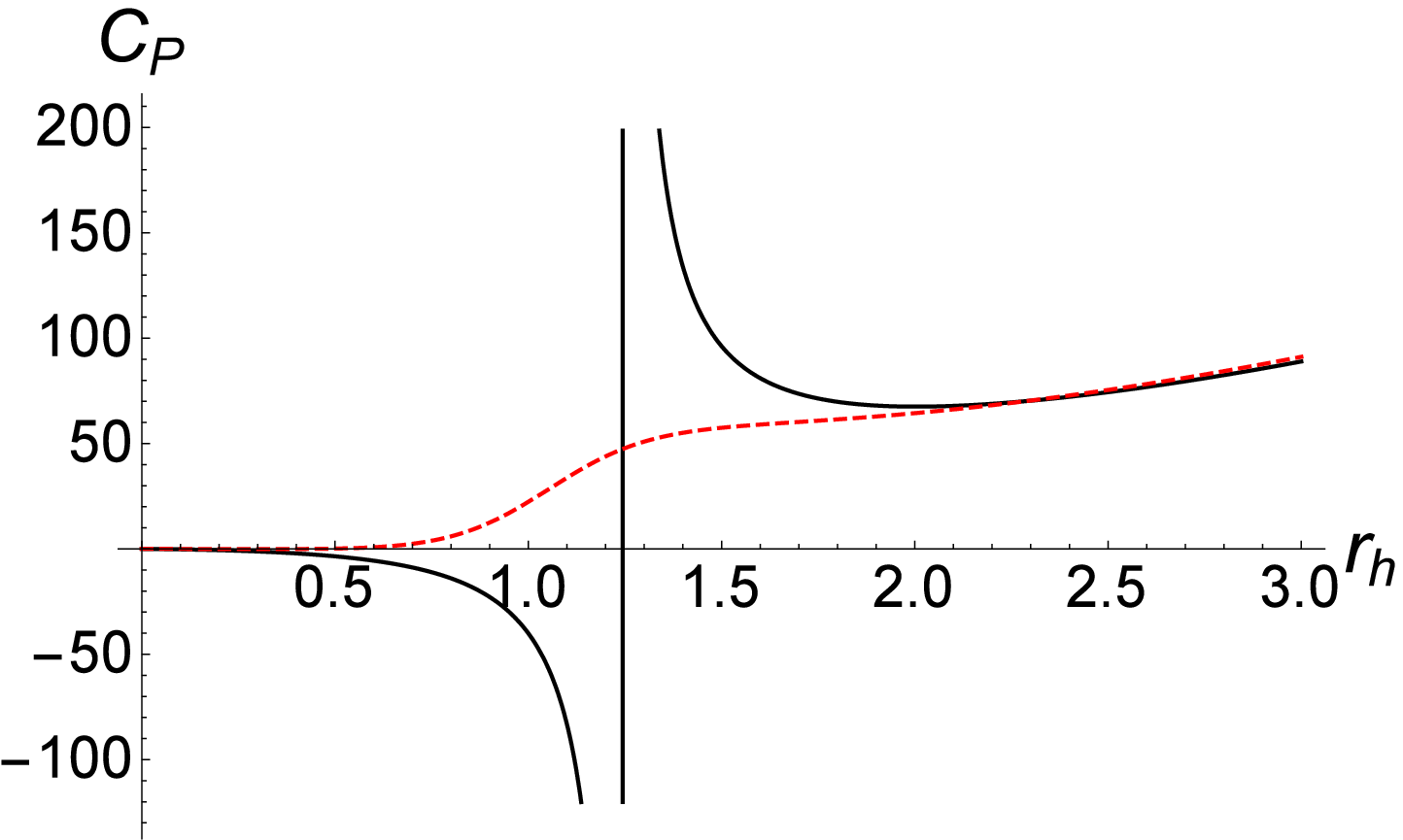}
	\caption{(Colour online). Semi-classical (black solid line) and GUP-corrected (red dashed line) heat capacities as function of $r_{h}$. We set (a) $a=\frac{1}{2\pi}$, $b=l=1$. (b) $b=0.1$, $a=l=1$.}
	\label{f4}
\end{figure}

Now, we investigate the behaviour of heat capacity for thermodynamic stability and possible phase transition of vdW black hole in Fig.(\ref{f4}). Again, both semi-classical and corrected heat capacities may show stable-unstable phase transitions. In Fig.(\ref{f4}a), we observe that corrected heat capacity diverges for $a=1/2\pi$ and $b=l=1$. So it shows the stable-unstable black hole  phase transition in the presence of GUP modification. Pradhan also reported a similar phase transition for vdW black hole in \cite{Pradhan2016}, but the phase transition occurs at the negative values of event horizon for $b=l=1$ and $a=1/2\pi$. Therefore, it is not physically acceptable. In Fig.(\ref{f4}b), both heat capacities change their roles for $a=l=1$ and $b=0.1$. While semi-classical heat capacity shows a stable-unstable phase transition, corrected heat capacity does not show any phase transitions.

\begin{figure}
	\centerline{\includegraphics[width=10cm]{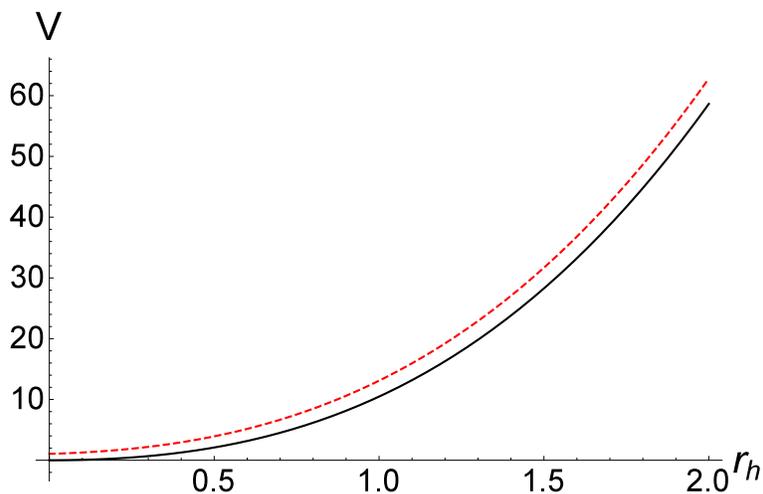}}
	\vspace*{8pt}
	\caption{Semi-classical (black solid line) and GUP-corrected (red dashed line) thermodynamic volume as function of $r_{h}$.  We set $b=1$.\label{f5}}
\end{figure}

\begin{figure}
	\centering
	\includegraphics[width=0.5\linewidth]{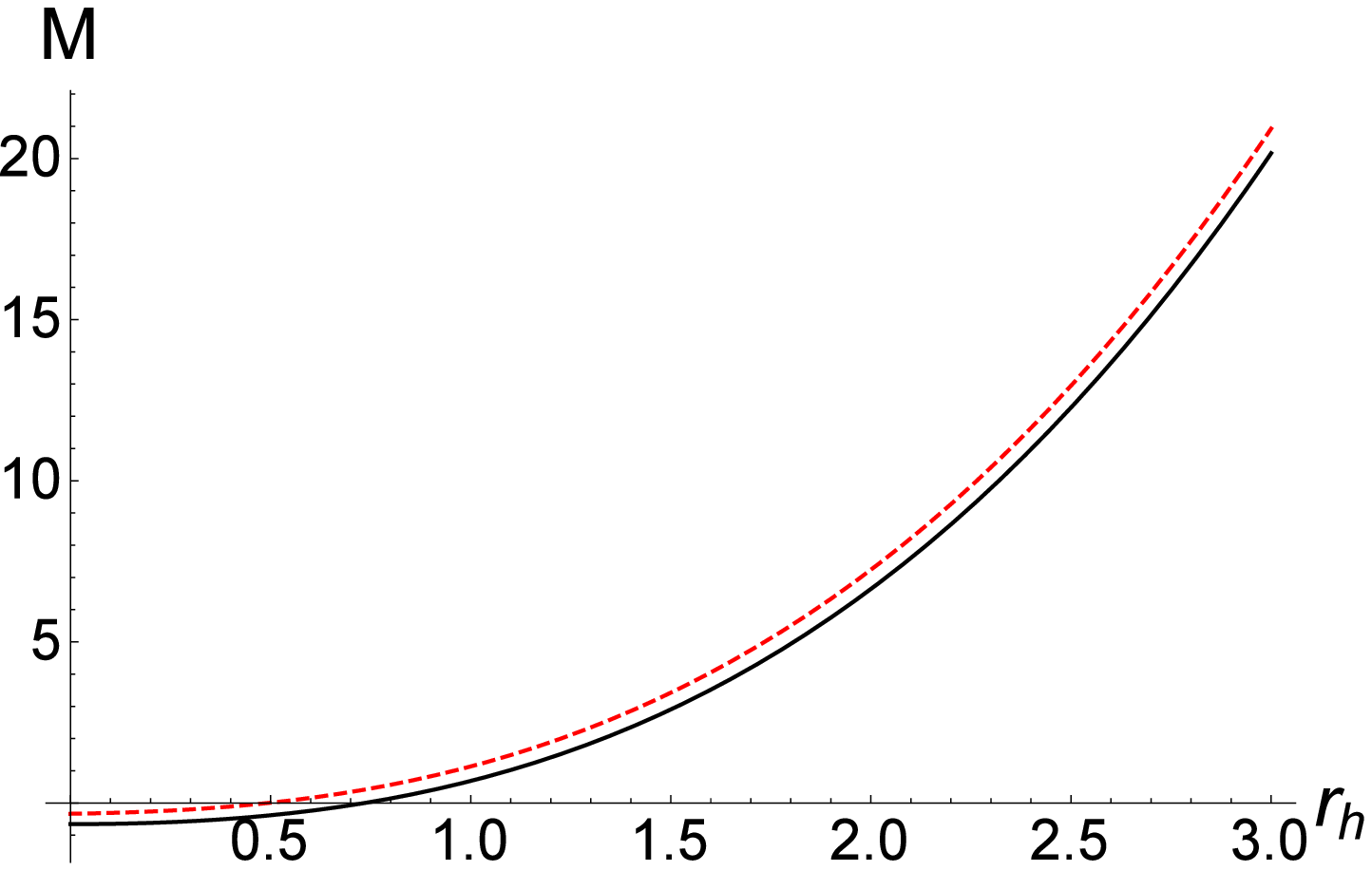}\hfil
	\includegraphics[width=0.5\linewidth]{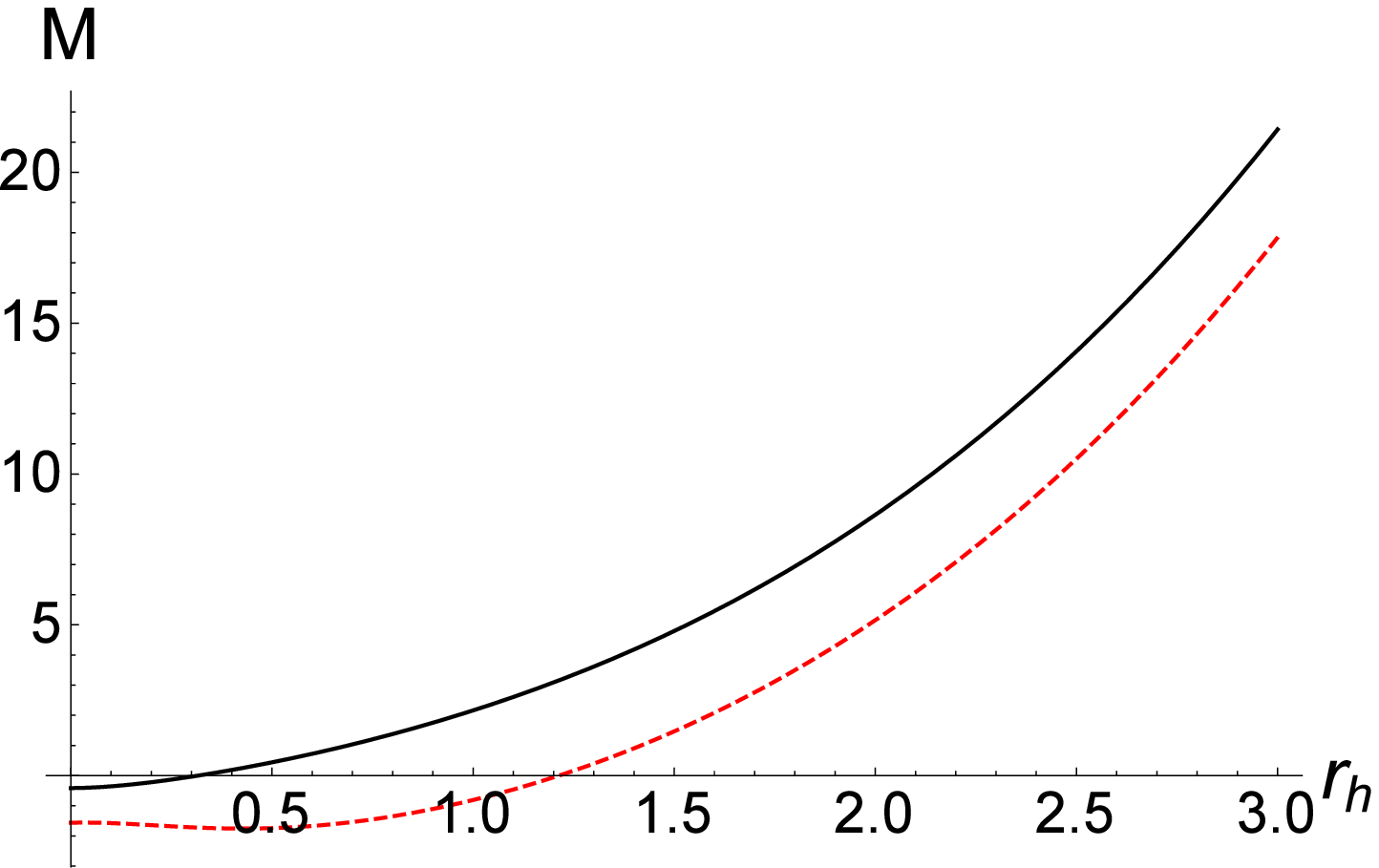}
	\caption{(Colour online). Semi-classical (black solid line) and GUP-corrected (red dashed line) masses as function of $r_{h}$. We set (a) $a=\frac{1}{2\pi}$, $b=l=1$. (b) $b=0.1$, $a=l=1$.}
	\label{f6}
\end{figure}

We investigate the behaviours of thermodynamic volumes and masses in Figs. (\ref{f5}) and (\ref{f6}), respectively. From (\ref{volume}), volume is given by
\begin{equation}
\label{volume2}
V=\frac{4}{3}\pi r_{h}^{3}+2\pi br_{h}^{2}+\frac{\pi b\alpha}{3}+\frac{\pi\alpha r_{h}}{2}\,.
\end{equation}
Since correction terms are always positive, modified volume is always larger than original volume. As it can be seen from Fig(\ref{f6}), corrected mass may be larger or smaller for a suitable choice of $a$ and $b$.
\begin{figure}
	\centering
	\includegraphics[width=0.5\linewidth]{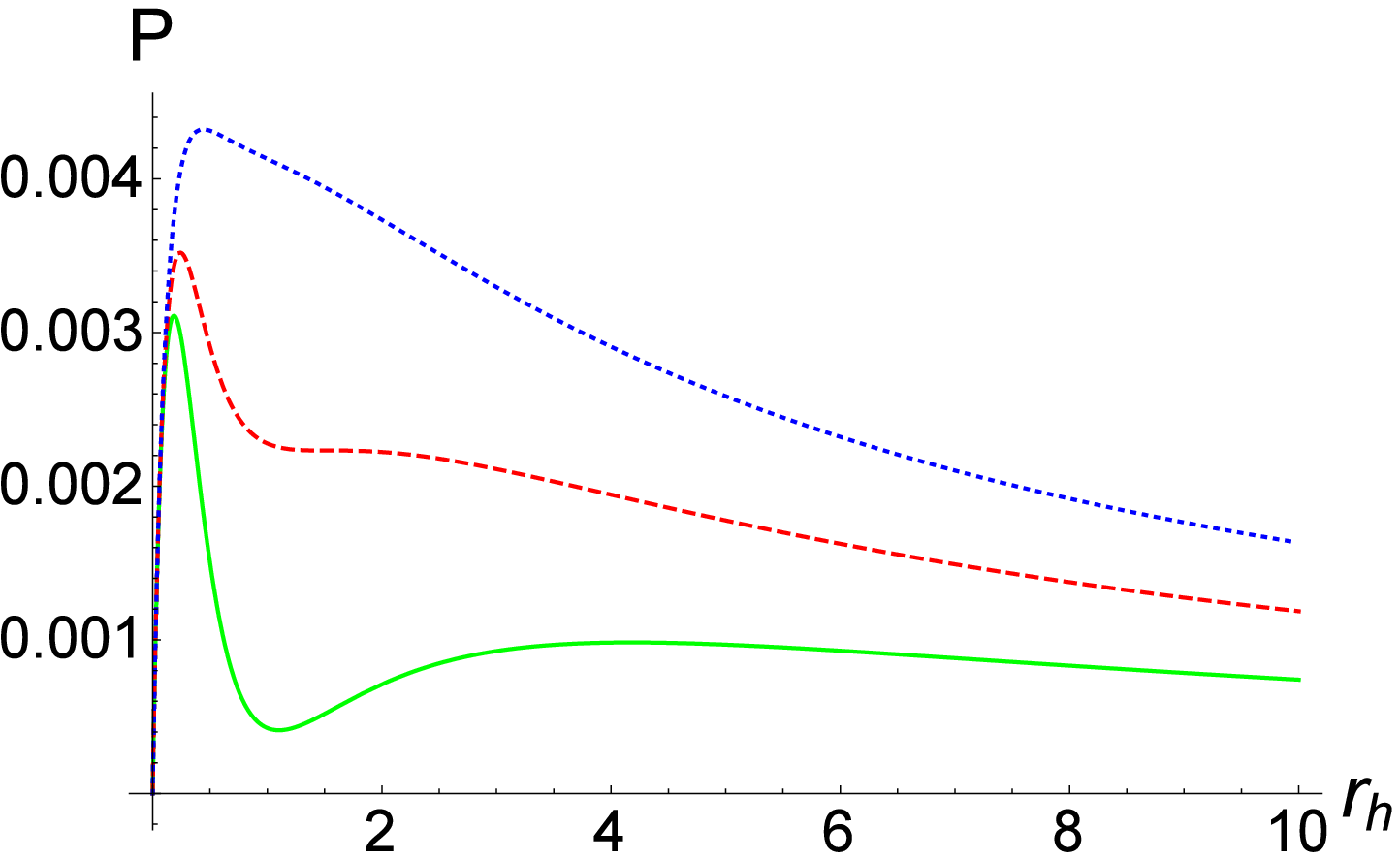}\hfil
	\includegraphics[width=0.5\linewidth]{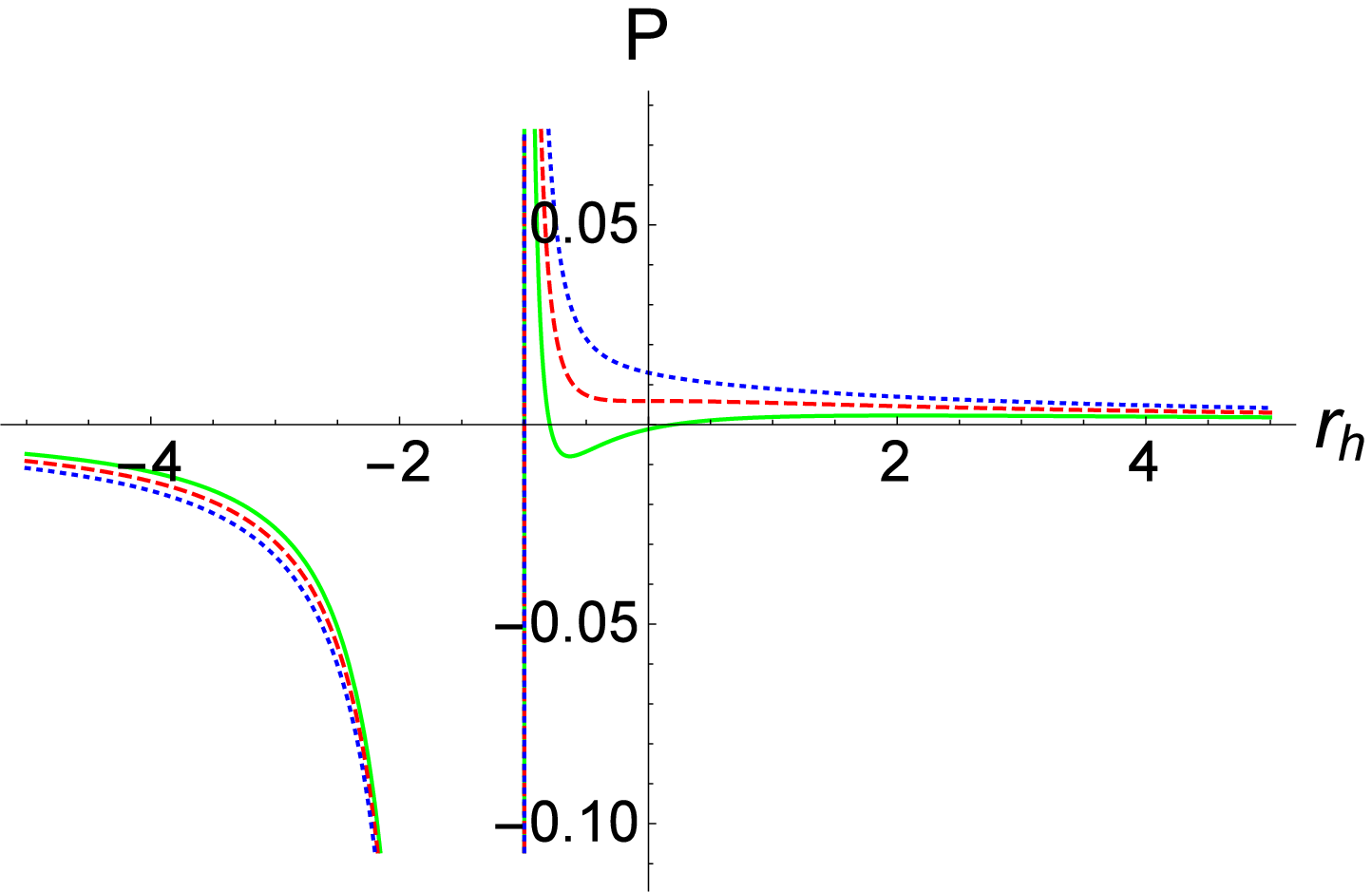}
	\caption{(Colour online). The $P-r_{h}$ diagram. The temperature of isotherms decrease from top to bottom and correspond to $1.3T_{c}$ (blue dotted line), $T_{c}$(red dashed line) and $0.7T_{c}$ (green solid line). We set $a=\frac{1}{2\pi}$, $b=1$. (a) GUP-corrected equation of state. (b) Original equation of state.}
	\label{f7}
\end{figure}

Needles to say, it is not a surprise to expect critical points since the black hole solution has the same thermodynamics with vdW fluids. From (\ref{temperature}) and corrected $h(r,P)$ function, one can obtain the equation of state
\begin{equation}
\label{Eos}
P=\frac{16r_{h}^{3}(2r_{h}+3b)((2r_{h}+3b)^{2}T-2a(r_{h}+b))+4ar_{h}(2r_{h}+b)(3r_{h}+2b)\alpha}{(2r_{h}+3b)^{3}(4r_{h}^{2}+\alpha)(8r_{h}(r_{h}+b)+\alpha)}\,.
\end{equation}
In order to obtain the critical points, we should solve the following two equations:
\begin{equation}
\frac{\partial P}{\partial r_{h}}=\frac{\partial^{2}P}{\partial r_{h}^{2}}=0\,.
\label{CP}
\end{equation}
Since the (\ref{Eos}) is very complicated, we have to find critical points numerically. We set $\alpha=b=1$ and $a=1/2\pi$. So we find $T_{c}=0.03279$, $r_{c}=1.52439$ and $P_{c}=0.00223$. In Fig.(\ref{f7}a), we show the small-large black hole phase transition for modified equation of state. Except the unstable left branch, the phase diagram mostly shows the well-known behaviour of liquid-gas system. Similar behaviours  for the GUP corrected phase transition of charged AdS black hole were also showed in the paper\cite{Sun2018} by Sun and Ma. Although original vdW black hole equation of state shows critical points, the phase transition seems physically unreasonable. One can obtain critical points from (\ref{CP}) 
\begin{equation}
T_{c}=\frac{8a}{27b},\qquad r_{c}=0\,\qquad P_{c}=\frac{a}{27b^{2}}\,.
\label{CP2}
\end{equation}
It is possible to define a positive critical specific volume, but small black hole  and phase transition branches correspond to negative event horizons. Therefore, the critical points do not correspond to physical phase transition in the absence of GUP modification.

\section{Conclusions}
\label{Concl.}
In this paper, we have considered the GUP-correction and obtained the GUP-corrected vdW black hole solution. We have also investigated the thermodynamic quantities and the phase transition of that black hole. Black hole thermodynamics should be modified since the quantum gravity effects are taken into account near the Planck scale. Therefore we have considered GUP modification for the vdW black holes. In order to obtain the solution, we have used the GUP-corrected black hole equation of state with that of the vdW fluids. Then, we have found the modified $h(r_{h},P)$ function.

Moreover, we have found the modified thermodynamic quantities such as entropy, temperature, heat capacity and thermodynamic volume. We have shown that GUP affects the thermodynamic behaviours of vdW black holes. GUP-corrected entropy is always smaller than the semi-classical entropy. Modified temperature may be larger or smaller than semi-classical temperature for a suitable choice of metric parameters. Corrected mass may also be larger or smaller than original mass. We have also found corrected volume is always bigger than semi-classical volume. In order to investigate any phase transitions and thermodynamical stabilities, we have analysed the specific heat. Again, one may observe a stable-unstable phase transition for both modified and semi-classical heat capacities. Finally, we have presented the P-V criticality of GUP-corrected vdW black holes. Although one finds critical points for semi-classical equation of state, the phase transition is not physically reasonable due to negative event horizons. 

In a summary, we have revealed some properties of thermodynamics and phase transition of vdW black holes due to the GUP-corrections. It is also interesting to study higher-dimensional vdW black holes for the GUP modifications since the importance of the particular dimensions\cite{Delsate2015} and our study can be generalized for the black hole solutions \cite{Setare2015,Abchouyeh2017,Xu2018,Debnath2018} in this direction.

\section*{Acknowledgement}
The authors thank the anonymous reviewers for their helpful and constructive comments.

\section*{Appendix}
In this appendix, we will give the field equations. Once the field equations are obtained, the corresponding stress energy tensor can be constructed from these equations. From (\ref{metric1}), the components of Ricci tensor $R_{\mu\nu}$ are given by
\begin{eqnarray}
\label{a1}
R_{00}=f\left(\frac{f''}{2}+\frac{f'}{r}\right) ,\qquad R_{11}=-\frac{1}{f}\left(\frac{f''}{2}+\frac{f'}{r}\right)\,, 
\end{eqnarray}
\begin{eqnarray}
\label{a2}
R_{22}=1-f'r-f ,\qquad R_{33}=R_{\theta\theta}\sin^{2}\theta\,,
\end{eqnarray}
and Ricci scalar R is given by
\begin{equation}
\label{a3}
R=-f''-\frac{4f'}{r}-\frac{2f}{r^{2}}+\frac{2}{r^{2}}\,.
\end{equation}
From $R_{\mu\nu}-\frac{1}{2}g_{\mu\nu}R+g_{\mu\nu}\Lambda=8\pi T_{\mu\nu}$, we get the components of stress energy tensor,
\begin{eqnarray}
\label{a4}
T_{00}=\frac{f-f^{2}-ff'r}{8\pi r^{2}}+Pf,\qquad T_{11}=\frac{f+f'r-1}{8\pi fr^{2}}-\frac{P}{f}\,,
\end{eqnarray}
\begin{eqnarray}
\label{a5}
T_{22}=\frac{2f'r+f''r^{2}}{16\pi}-Pr^{2},\qquad T_{33}=T_{22}\sin^{2}\theta\,,
\end{eqnarray}
where we use $\Lambda=-8\pi P$. Taking $T^{\mu\nu}$ as an anisotropic fluid source in (\ref{stressEnergy}), one can easily find (\ref{energyDensityWithp1}) and (\ref{principalPressure23}).

\end{document}